\documentclass{article}
\usepackage{spconf,amsmath,graphicx,hyperref}
\usepackage{booktabs}
\usepackage{arydshln}

\title{Zero-Shot TTS with Enhanced Audio Prompts: BSC submission for the 2026 Wildspoof challenge TTS track \thanks{ \copyright 2026 IEEE. Personal use of this material is permitted. Permission from IEEE must be obtained for all other uses, in any current or future media, including reprinting/republishing this material for advertising or promotional purposes, creating new collective works, for resale or redistribution to servers or lists, or reuse of any copyrighted component of this work in other works.}}
\name{Jose Giraldo$^{1}$, Alex Peir\'{o}-Lilja$^{1,2}$, Rodolfo Zevallos$^{1}$, Cristina España-Bonet$^{1,3}$}
\address{$^{1}$Langtech Lab, Barcelona Supercomputing Center, Catalonia, Spain;  
$^{2}$Centre de Llenguatge i Computació, \\ Universitat de Barcelona, Spain; 
$^{3}$DFKI GmbH, Saarland Informatics Campus, Germany}
\begin{document}
%
\maketitle
\begin{abstract}
We evaluate two non-autoregressive architectures, StyleTTS2 and F5-TTS, to address the spontaneous nature of in-the-wild speech. Our models utilize flexible duration modeling to improve prosodic naturalness. To handle acoustic noise, we implement a multi-stage enhancement pipeline using the Sidon model, which significantly outperforms standard Demucs in signal quality. Experimental results show that finetuning enhanced audios yields superior robustness, achieving up to 4.21 UTMOS and 3.47 DNSMOS. Furthermore, we analyze the impact of reference prompt quality and length on zero-shot synthesis performance, demonstrating the effectiveness of our approach for realistic speech generation.
\end{abstract}

\section{Introduction}
\label{sec:intro}


Traditional Text-to-speech (TTS) systems rely on high quality studio data that usually comes from the audiobook domain. 
LibriTTS \cite{Zen2019LibriTTSAC}, MLS \cite{Pratap2020MLSAL}, and many more are examples of this usage. However, there is a growing trend for expressive TTS models that are able to synthesize spontaneous speech, needing data that is captured in the wild. Emilia~\cite{he2024emiliaextensivemultilingualdiverse} along with The \emph{TTS In the Wild} (TITW) dataset~\cite{jung25c_interspeech} provide spontaneous speech data to train TTS systems; the latter is designed to foster research in the training of TTS systems using noisier datasets.

Training TTS models on in-the-wild data presents several challenges. First, the variety of environmental noise and acoustic conditions can introduce errors in automatic transcriptions, which can prevent convergence to intelligible speech for many architectures. Second, the diversity in prosody and speaking styles including hesitations, fillers, and variable pacing makes duration modeling substantially more difficult than in controlled audiobook recordings.
%

To address these challenges, our approach combines: 
(1) speech enhancement to improve training data quality, 
(2) non-autoregressive architectures (F5-TTS \cite{chen2025f5} and StyleTTS2 \cite{li2023styletts}) with flexible duration modeling, and 
(3) systematic analysis of inference parameters (prompt quality and length) to maximize audio quality and intelligibility.

\section{Experiments}
\label{sec:experiments}

Given the spontaneous nature of the dataset and the zero-shot generation for one of the evaluation protocols StyleTTS2 and F5-TTS were chosen.
We initially experimented with training from scratch a smaller variant of F5-TTS (12 layers, 16 heads) using the findings of \cite{Saratchandran2025LeanerTM}. However, this approach underperformed compared to finetuning strategies. Both models were finetuned on the \emph{TITW Easy partition} after an enhancement step with the Sidon model~\cite{nakata2025sidon}. Although TITW-Easy already had an enhancement step with Demucs~\cite{défossez2019demucsdeepextractormusic}, Sidon can increase the Mean opinion score (MOS) compared to Demucs by a large margin. 

F5-TTS was finetuned from the pretrained \emph{\small F5TTS\_v1\_Base}\footnote{\url{hf.co/SWivid/F5-TTS/tree/main/F5TTS_v1_Base}} checkpoint for 75,000 steps using a learning rate of 1e-5, with a warm-up phase for the initial 5,000 updates. Batch size was set to 76,800 audio frames. To accelerate convergence and maintain consistency with the pretrained model's text representation space, we did not change the tokenizer vocabulary and use the same as the pretrained model. 

StyleTTS2 was finetuned from the LibriTTS pretrained checkpoint\footnote{\url{hf.co/yl4579/StyleTTS2-LibriTTS} } using a learning rate of 1e-4, batch size of 16 and maximum length of 800 for 12,000 steps (5 epochs), diffusion training was set after the first iteration and joint training started at the second epoch. 

At inference time, we also explored the impact of the quality and length of the reference audio, since both models required it to perform inference. We used VERSA \cite{shi2025versaversatileevaluationtoolkit} to compute the evaluation metrics in different dimensions. For audio quality: UTMOS \cite{saeki2022utmosutokyosarulabvoicemoschallenge} and DNSMOS pro \cite{cumlin24_interspeech}, for intelligibility: the Word Error Rate (WER) of the synthesized speech transcription using Nemo conformer large model compared to the input text,  and for speaker similarity F0 RMSE and the Speaker Encoder Cosine Similarity (SECS) using the embeddings of the model \emph{voxcelebs12\_rawnet3} from ESPnet~\cite{jung2024espnet}.

\section{Results}
\label{sec:results}

As a first experiment, the length of the audio prompt was chosen between the minimum (\emph{short}) and maximum (\emph{long}) audio available from each speaker in the TITW-KSKT subset (Known Speaker, Known Text). Although, the average of the audio prompt duration between long and short prompts differs only by two seconds, the speaker similarity drops in both models when using the short prompt as seen in Table \ref{tab:prompt-length}. For StyleTTS2 the degradation is more pronounced, the WER rises up to 0.49 in contrast with F5-TTS where the WER is unaffected. Our findings suggest a direct correlation between prompt length and speaker similarity, corroborating prior work in zero-shot TTS \cite{Lei2023ImprovingLM}.

\begin{table}[t]
\centering
\caption{Effect of audio prompt length on model performance (on the KSKT set)}
\label{tab:prompt-length}
\resizebox{\columnwidth}{!}{
\begin{tabular}{l c c c c c}
\toprule
      & Duration &  UTMOS $\uparrow$ & SECS $\uparrow$ & F0 RMSE $\downarrow$ & WER $\downarrow$ \\ 
Model &  Avg (s) & &  & \\ 
\midrule
F5-TTS$_{long}$ & 7.7     & 3.51 & \textbf{0.35} & 51.49 & \textbf{0.07} \\
F5-TTS$_{short}$ & 5.5     & \textbf{3.62} & 0.24 & 54.71 & 0.07 \\
\midrule
StyleTTS2$_{long}$ & 7.7  & 3.37 & 0.19 & \textbf{51.08} & 0.21 \\
StyleTTS2$_{short}$ & 5.5  & 2.56 & 0.14 & 55.30 & 0.49 \\
\bottomrule
\end{tabular}
}
\end{table}

We were also interested in the effect of the audio quality of the audio prompt. Building on the results from the previous experiment we use the longest audio available per speaker with and without enhancement to perform the inference in both sets (KSKT and KSUT). As observed in Table \ref{tab:audio-quality}, using the enhanced audio prompt improves the audio quality scores (UTMOS and DNSMOS) in both models, however, the speaker similarity scores are degraded. The WER is also improved in both models when using the enhanced reference, but the effect is more noticeable in StyleTTS2. This could be due to the  different role of reference audio in each model, in StyleTTS2 the reference audio is the input to an acoustic encoder while on F5-TTS the reference audio is treated as a prompt that the model uses to fill the masked region conditioned on the input text. 

For comparison, we include results for the smaller F5-TTS variant (12 layers, 16 heads) trained from scratch for 1M steps on TITW (F5-TTS{\footnotesize $tiny$} in Table \ref{tab:audio-quality}). This approach achieved only 3.27 UTMOS and poor speaker similarity (0.10), confirming the value of transfer learning from large-scale speech pretraining. In contrast to Hierspeech++~\cite{lee2023hierspeechbridginggapsemantic} results in denoising speech prompts, we did not find a degradation in intelligibility when using enhanced reference audios; this is due to their usage of MP-SENet as enhancement model which is known to have content preservation issues, but Sidon is able to deliver high quality audios while preserving the message.

\begin{table}[h]
\centering
\caption{Effect of enhancement in the audio prompt (\emph{Enh.}) on model performance. All models use the long prompt strategy}
\label{tab:audio-quality}
\resizebox{\columnwidth}{!}{
\begin{tabular}{l l c c c c c}
\toprule
 & Test & UTMOS $\uparrow$ & DNSMOS $\uparrow$ & WER $\downarrow$  & SECS $\uparrow$ & F0 RMSE $\downarrow$ \\ 
Model &  & & (pro bvcc) & & \\ 
\midrule
F5-TTS$tiny$ & KSKT & 3.27 & 2.57 & 0.20 & 0.10 & 48.41 \\
\hdashline
F5-TTS        & KSKT & 3.51 & 3.11 & 0.08 & \textbf{0.35} & 51.49 \\
\quad(+Enh. prompt)    & KSKT & 3.89 & \textbf{3.31} & \textbf{0.07} & 0.28 & 52.05 \\
\hdashline
StyleTTS2 & KSKT & 3.37 & 2.34 & 0.21 & 0.19 & 51.08 \\
\quad(+Enh. prompt)  & KSKT & \textbf{3.97} & 2.82 & 0.14 & 0.18 & \textbf{45.85} \\
\midrule
F5-TTS        & KSUT & 3.68 & 3.22 & 0.14 & - & - \\
\quad(+Enh. prompt)      & KSUT & 4.02 & \textbf{3.47} & 0.13 & - & - \\
\hdashline
StyleTTS2 & KSUT & 3.59 & 2.55 & 0.18 & - & - \\
\quad(+Enh. prompt) & KSUT & \textbf{4.21} & 2.99 & \textbf{0.10} & - & - \\
\bottomrule
\end{tabular}
}
\end{table}



\begin{figure}[h]
\begin{minipage}[b]{1.0\linewidth}
  \centerline{\includegraphics[width=8.1cm]{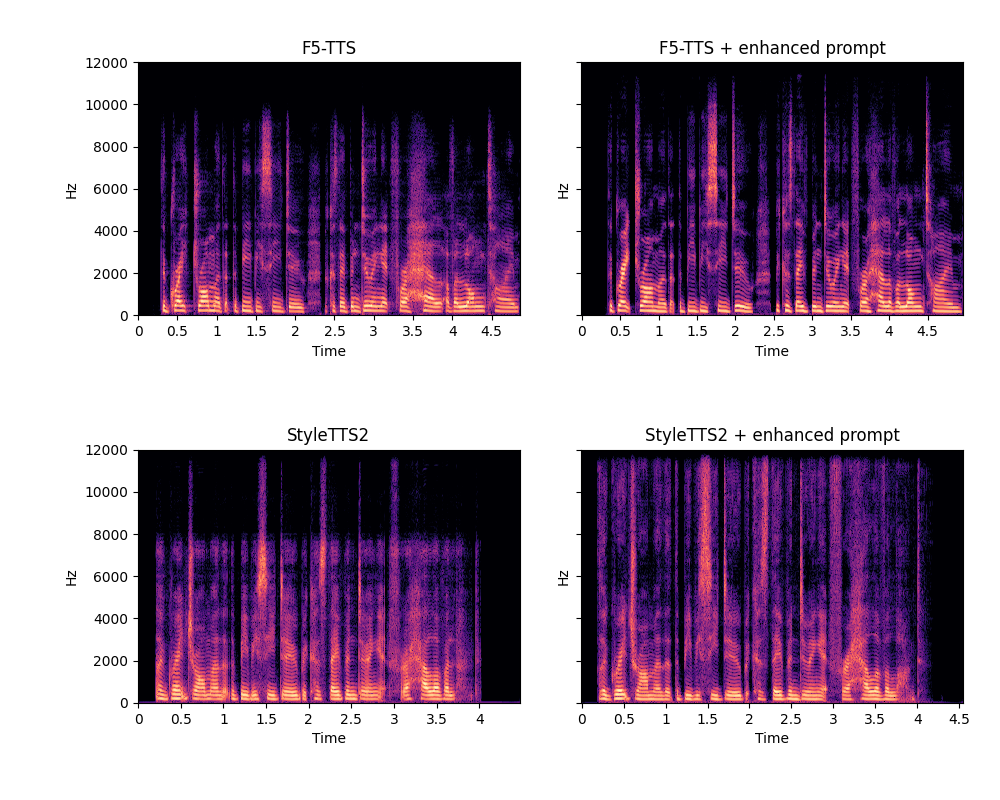}}
\vspace{-1.5em}
\end{minipage}
\caption{Spectrogram comparison of models with different input audio prompts}
\label{fig:res}
\end{figure}

Visual inspection of the spectrograms (Figure \ref{fig:res}) reveals distinct enhancement effects for each model. F5-TTS without enhancement produces limited bandwidth, with minimal energy above 8 kHz; enhancement extends the bandwidth and recovers high-frequency harmonics with greater clarity. StyleTTS2 without enhancement already generates some content above 8 kHz, but enhancement produces more consistent high-frequency energy across the entire utterance, resulting in fuller spectral content.

\section{Conclusion and future work}

\label{sec:conclusion}

For the WildSpoof challenge, we selected the samples generated with F5-TTS with enhanced reference prompts for the final submission. This was our best model w.r.t intelligibility (WER), audio quality (DNSMOS pro) and speaker similarity. The length and quality of the audio prompt were a key factor to maximize the performance of  the benchmark samples in the evaluation metrics. Future work should investigate adaptive enhancement strategies that balance audio quality with speaker identity preservation.

\section{Acknowledgements}

This work has been funded by the Government of Catalonia through the Aina project and by the MTDFP ministry and Plan de Recuperación, Transformación y Resiliencia (PRTR) ---Funded by EU--- NextGenerationEU within the framework of the project Desarrollo Modelos ALIA. CEB acknowledges her AI4S fellowship within the “Generación D” initiative by Red.es, MTDFP, for talent attraction (C005/24-ED CV1), funded by NextGenerationEU through PRTR.

\bibliographystyle{IEEEbib}
\bibliography{strings,refs}

\end{document}